\documentclass[twocolumn,prd,aps,showpacs,reprint]{revtex4-1} 
\usepackage{amsmath,mathrsfs,amsbsy,color,graphicx,bm,amsthm,amsfonts}
\usepackage{mathtools}
\usepackage{times}
\usepackage{graphicx}
\usepackage{ulem}
\usepackage{braket}
\usepackage{dsfont}

\newcommand{\be}{\begin{equation}}
\newcommand{\ee}{\end{equation}}
\newcommand{\br}{\begin{eqnarray}}
\newcommand{\er}{\end{eqnarray}}

\newtheorem{lemm}{Lemma}

\begin{document}
\title{Non-linear equation of motion for Page-Wootters mechanism with interaction and quasi-ideal clocks}
\author{Leandro R. S. Mendes } 
\email{leandrorsm@gmail.com}
\author{Frederico Brito}
\email{fbb@ifsc.usp.br}
\author{Diogo O. Soares-Pinto}
\email{dosp@usp.br}
\affiliation{Instituto de F\'isica de S\~ao Carlos, Universidade de S\~ao Paulo, CP 369, 13560-970, S\~ao Carlos, S\~ao Paulo, Brazil}

\begin{abstract}
Among the many proposals to approach the concept of time in quantum theory, the Page-Wootters mechanism has attracted much attention in the last few years. Originally, such a mechanism explored a stationary bipartite non-interacting global system, i.e., a system of interest together with an ancillary clock, to determine how the evolution in time can emerge as an equation of motion for a quantum particle conditioned to the measurement of the state of the clock. After the seminal proposal, many variations of it were considered, and different aspects of the mechanism were elucidated. Our contribution to these investigations is that we analyze such a timeless approach to quantum theory but deriving an equation of motion for a mixed state system that evolves according to its gravitationally induced interaction with a non-ideal quantum clock. The interaction considered is known to describe the gravitational decoherence mechanism, and the clock model is the recently proposed quasi-ideal clock, i.e., one constructed to approximate the time-energy canonical commutation relation. As a result of our considerations, we obtained an equation of motion that is non-linear in nature, dependent on the system's initial conditions.  

\end{abstract}

\maketitle

\section{Introduction}

The goal to unify quantum theory and Einstein's theory of gravitation has demonstrated itself to be an arduous task. Meanwhile, to not having a theory of quantum gravity, much effort is being put into trying to understand scenarios where we can study the effects of relativity on quantum mechanical systems without the commitment to high energies \cite{wood1,smith1,bose,chia}. One way to achieve this is to take lessons from general relativity and investigate constructions for the quantum theory which do not depend on background structures, such as time. In quantum mechanics, time has a special place; it is a parameter that in respect to which all other events happen, a preference that is certainly not exhibited by general relativity, as a result of the principle of general covariance \cite{carlo}. 

Time covariant quantum mechanics, where the dependence of the states on an external time reference frame is excluded and therefore time is treated like any other quantity, can exhibit a Hamiltonian constraint that takes the form of a Wheeler-DeWitt equation, rendering a timeless condition. The dynamics in a timeless formalism can be recovered by assigning a clock, as the state of one of the subsystems, regarding to which time is referred. Then in this construction, time is what is read by a clock, in the same manner as viewed in general relativity. This way to restore a kind of time evolution in a static global state is recognized as the Page-Wootters formulation \cite{paw,f3} also known as the timeless approach to quantum mechanics and relational Schr{\"o}dinger picture, an approach that is gathering much attention \cite{diaz,foti,gamb1,hoen,pegg,f,f1,f2,f4,f5}.  

Many advances were achieved in this formulation. Among them, they accomplished to resolve criticisms about the propagators of the formulation \cite{glm}, showed causal relation on the change in reference frames \cite{casl}, established limits on the precision on constructing an observable to measure time \cite{smith3}. A point in common for all the advances mentioned above is the use of idealized quantum clocks. Even though such clocks are usually sufficient to extract the physics of the problem, we can wonder what would be the case for a more realistic scenario where we rely on a non-ideal clock, i.e., a system with finite dimension and states which are not necessarily orthogonal. Although there are several models proposed for the construction of quantum clocks, this work is especially interested in a clock model that mimics several properties of the ideal clock while having finite dimension. Such a clock is known as a \textit{quasi-ideal clock} \cite{mischa}, which was already shown to be able to satisfy a covariant condition akin to the Page-Wootters condition \cite{cont}.

Here, we further expand the timeless formulation of quantum mechanics in a two-fold way: we employ the quasi-ideal clock model as the ``timekeeper" in the timeless formulation obtaining a Schr{\"o}dinger-like equation for the case where clock and system interact gravitationally. In doing that, we extend a previous result where it was obtained a time non-local Schr{\"o}dinger equation when the system is in a pure state, and there is an interaction \cite{smith2}. Additionally, we consider the system to be in a mixed state, and we derive an equation of motion that describes gravitational decoherence. This equation is unique in that it depends on the system's initial conditions, therefore, being non-linear in nature. The fact that the clock is not ideal is reflected in a dependence upon its dimension, where the dimension alleviates these non-linear and non-unitary effects.

\section{Time non-local Schr{\"o}dinger equation}

We start by briefly revising previous work \cite{paw, smith2} describing the effects of interaction in the Page-Wootters model. It is considered a physical Hilbert space $\mathcal{H}_{phys}$ composed by all global states $\ket{\Psi}$, which satisfy a Wheeler-DeWitt-like constraint \cite{whe,wit} \be\label{fis} H\ket{\Psi}=0, \ee where $H$ is the total Hamiltonian. The global state is considered to comprise a clock and a system, being the clock the timekeeper of the system.  Defining an initial clock state $\ket{0}$ and demanding state evolution between these states to be \be\label{ev} \ket{\tau} = e^{-i H_{C} \tau }\ket{0},\ee any state of the system can be obtained by conditioning $\ket{\Psi}$ to a clock state associated to a time $\tau$, as \be\label{um} \ket{\psi_{S}(\tau)} = \left(\bra{\tau}\otimes \mathds{1}_{s}\right) \ket{\Psi},\ee where $\ket{\psi_{S}(\tau)}$ is a system state and $\ket{\tau}$ is the clock state associated with time $\tau$. From Eqs. (\ref{fis}) and (\ref{um}) we obtain an equation of motion for the system state \be i\frac{d}{d\tau}\ket{\psi_{S}(\tau)} = H_S\ket{\psi_{S}(\tau)},\ee a Schr{\"o}dinger-like equation with respect to the clock parameter $\tau$.  When an interaction between the clock and the system is taken into account, the constraint above in Eq.(\ref{fis}), can be read as \be\label{tw}  (H_S+H_C+H_I)\ket{\Psi}=0, \ee where $H_S$, $H_C$ and $H_I$ are the system, clock and interaction Hamiltonians, respectively. Using Eq.(\ref{tw}), we can still have a Schr{\"o}dinger-like evolution, more specifically the system state will obey a time non-local Schr{\"o}dinger equation \be \label{amd} i\frac{d}{d\tau}\ket{\psi_{S}(\tau)} = H_S \ket{\psi_S(\tau)} + \int{d\tau' K(\tau,\tau')\ket{\psi_{S}(\tau')}}, \ee were $K(\tau,\tau')=\bra{\tau}H_{I}\ket{\tau'}$ is seen as the kernel connected to the interaction Hamiltonian. Given that this equation is non-local in time, it is implied that to verify the solution for this equation it is required knowledge of the system state at all times \cite{smith2}.

\section{Ideal and Quasi-Ideal clocks}

The result obtained in Eq.(\ref{amd}), as many others, employ the use of ideal clocks that in the context of the Page-Wootter's mechanism is usually considered to be infinite-dimensional, possessing a distinguishable basis of time states and being associated to a time observable $\mathcal{T}$. The use of an ideal clock can be traced back to the requirement that time observable will obey the canonical commutation relation with the system Hamiltonian $H$ \be [H,\mathcal{T}] = -i, \ee being equivalent to demanding that the expectation value of $\mathcal{T}$ will vary linearly with time. This condition is, supposedly, only obeyed by the ideal momentum clock \cite{pash}, with clock Hamiltonian $H_C=P_C$ the momentum operator and position operator as the time observable $X=\mathcal{T}$.  

Respecting this condition is not trivial, and well-known clock models fail to adhere to it. Notably, even the prominent Salecker-Wigner-Peres (SWP) clock cannot approximately approach such condition \cite{sw,swp}. The discussion surrounding the construction of a time observable is vast, and we will not delve further into it. What is of interest in the present work is that there are constructions that can approximate this relation of canonical commutation. More specifically, we focus on the construction of the quasi-ideal clock (QIC), for which \be [H,\mathcal{T}]\ket{\psi_{QI}(k_0)} = -i\ket{\psi_{QI}(k_0)} + \ket{\varepsilon_{comm}}, \ee being $\ket{\psi_{QI}(k_0)}$ the QIC state (which will be presented shortly) and $\ket{\varepsilon_{comm}}$, a non-unity state vector which quantifies the error in the commutation relation of the QIC when compared to the ideal case, being exponentially small in dimension \cite{mischa}.

The QIC is based on the SWP clock, which can be seen as a quantum rotor \cite{roto} having a Hamiltonian with evenly spaced energy levels \be H_{C}=\sum^{d-1}_{j=0} \omega j \ket{E_j}\bra{E_j},\ee where $d$ is the dimension of its Hilbert space. The SWP clock states form an orthonormal set of states $\{\ket{\theta_{k}}\}$ for $k=0,1, \dots ,d-1$ which is dubbed the time basis. These states are connected to the energy eigenstates through a discrete Fourier transform \be \ket{\theta_k} = \frac{1}{\sqrt{d}}\sum^{d-1}_{j=0} e^{-i2\pi j k/d}\ket{E_j},\ee and evolve according to $e^{-iH_{C}T/d}\ket{\theta_{k}} = \ket{\theta_{k+1}}$, thus establishing regular, integer, time intervals $t_k= (T/d)k$, such that the periodic condition $e^{-iH_{C}T}\ket{\theta_{k}} = \ket{\theta_{\mod(k, d)}}$ is satisfied, and providing means to the construction of the time operator \be \mathcal{T} = \sum_{k} t_k \ket{\theta_k}\bra{\theta_k}.\ee

The QIC is defined as a coherent superposition of time states \cite{mischa} \be \ket{\psi_{QI}(k_{0})} = \sum_{k \in S_d(k_{0})} A e^{-\pi (k-k_{0})^{2}/\sigma^{2}} e^{i2\pi j_{0} (k-k_{0})/d}\ket{\theta_{k}}, \ee where $A$ is a normalization constant, $S_d(k_0)$ is a set of $d$ consecutive integers centered about $k_0$, which is considered the initial time position of the QIC, and $\omega j_0$ is the average energy of the clock. From here on, we are going to consider $\psi(k_{0}; k) = A e^{-\pi (k-k_{0})^{2}/\sigma^{2}} e^{i2\pi j_{0} (k-k_{0})/d}$. For this model of clock exists a relation equivalent to Eq.(\ref{ev}), \be \label{evol} e^{-iH_{C}t}\ket{\psi_{QI}(k_{0})} = \ket{\psi_{QI}\left(k_{0}+ td/T\right)} + \ket{\varepsilon}, \ee where $T$ is the period of the clock, $\ket{\varepsilon}$ is an error term composed of an error in changing the mean position of the clock, i.e., passing from $k_0$ to $k_0 +1$ and an error in changing the consecutive integers, i.e., $S(k_0)$ to $S(k_0 +1)$, whose norm is exponentially small with dimension. Differently from the SWP clock, the quasi-ideal states are continuous, in the sense that its evolution holds for arbitrarily small time intervals, where $t \in \mathds{R}$.

\section{Relative state for the Quasi Ideal clock}

From now on, without loss of generality, we choose the initial state of the clock to be centered around $k_0 = 0$ and rewrite the evolution, so that Eq.(\ref{evol}) become  \be  e^{-i\tau H_{C}}\ket{\psi_{QI}(0)} = \ket{\psi_{QI}\left(\tau d/T\right)} + \ket{\varepsilon}. \ee As in Eq.(\ref{um}) we wish to define the relative state of the system as being a slice of the global state in respect to the QIC at time $\tau$ \be \label{evert} \ket{\psi_{S}(\tau d/T)}_{e} = \braket{\psi_{QI}\left(\tau d/T\right)|\Psi}. \ee 
Although we are defining the system state with the parameter $\tau d/T$, from here on we are going to make the change $\tau d/T \rightarrow \tau$, so that all results will be written in terms of $\tau$. 

The subscript in Eq.(\ref{evert}) indicates that the state is an effective state for the system as can be seen in the equation below (see Appendix A)\br\label{sever} \ket{\psi_{S}(\tau)}_{e} &=& \frac{1}{T}\int^{T}_{0}d\tau' \braket{\psi_{QI}\left(\tau\right)|\psi_{QI}\left(\tau^{\prime}\right)}\ket{\psi_{S}(\tau')} \nonumber \\  &=& \frac{1}{T}\int^{T}_{0}d\tau' F_{QI}(\tau-\tau')\ket{\psi_{S}(\tau')}. \er The state we obtain is not exactly $\ket{\psi_{S}(\tau)}$, the one we would like to prepare, but an approximation that is dependent on how well the QIC can distinguish the clock states at different times. This effect is due to the non-ideal condition of the clock: time will be known with a certain accuracy that depends on the function $F_{QI}(\tau-\tau')$. In order for this function to become sharply peaked around $\tau$, and hence increasing the accuracy of the clock, we need to time-squeeze the QIC, which is accomplished for a regime where $\sigma < \sqrt{d}$. Thereby, we reduce the uncertainty in time readings in exchange for an increased uncertainty in energy, turning the QIC closer to a time state. This seems to come in detriment of quantum control \cite{mischa}: For squeezed states, the QIC becomes more vulnerable to the back-reaction of implementing unitaries, i.e., its errors do not decay exponentially with dimension any longer. This appears to imply that if it were to be implemented a quantum control in this mechanism, it would be necessary to introduce another clock for that purpose only, and any dynamics would be in reference to the time of the first clock.   
   
Another aspect of Eq.(\ref{evert}) worth noting is that we consider, here on, that the dimension of the clock is large enough that we can ignore the error on the evolution given by Eq.(\ref{evol}) considering $\braket{\psi_{QI}(\tau)|\Psi}\approx\bra{\psi_{QI}(0)}e^{i\tau H_C}\ket{\Psi}$. Thus, using the relative state definition we can obtain a slightly different form for the effective system state, which is going to be useful \br\label{def} \ket{\psi_{S}(\tau)}_e &=&  \sum_{k \in S_d(\tau)} \psi^{*}\left(\tau; k\right)\braket{\theta_{k}|\Psi} \nonumber \\  &=&  \frac{1}{\sqrt{d}}\sum_{k \in S_d(\tau)} \psi^{*}\left(\tau; k\right)\ket{\psi_{S}(k)}, \er where we used that $\ket{\psi_S(k)}= \sqrt{d}\braket{\theta_{k}|\Psi}$. We can see that, written in this way, the effective system state is a superposition of the system state conditioned to the time basis $\ket{\psi_{S}(k)}$. The relative state $\ket{\psi_{S}(k)}$ is not an effective version as in Eq.(\ref{evert}), due to the fact that the time basis form a complete distinguishable basis. Here on we are going to refer to the effective system state as only the system state, for simplicity.

\section{Equation of motion}

To obtain the equation of motion for the mixed-state system we first derive an equation of motion for the pure-state system with QIC. Starting from a lemma (see Appendix B) describing the time evolution for the QIC in respect to $\tau$ \begin{lemm}{Given a QIC state evolved to time $\tau$, its time evolution will be given by \be\label{sc} \frac{d}{d\tau}\ket{\psi_{QI}\left(\tau\right)} = - i\frac{T}{d}H_C\ket{\psi_{QI}\left(\tau\right)} - \ket{\varepsilon'},\ee} with error decreasing exponentially with dimension \be |\hspace{-0.05cm}|\ket{\varepsilon'}|\hspace{-0.05cm}| \leq \mathcal{O}(poly(d)e^{\frac{-\pi d}{4}}). \ee \end{lemm} We obtain a Schr{\"o}dinger-like equation for the system \br\label{seg} i\frac{d}{d\tau}\ket{\psi_{S}(\tau)}_e &=& i\frac{d}{d\tau}\braket{\psi_{QI}\left(\tau\right)|\Psi} \nonumber \\ &=& \left(-\frac{T}{d}\bra{\psi_{QI}\left(\tau\right)}H_{C} + i\bra{\varepsilon'} \right)\ket{\Psi} \nonumber \\ &=&  -\frac{T}{d}\bra{\psi_{QI}\left(\tau\right)}H_{C}\ket{\Psi} + i\braket{\varepsilon'|\Psi} \nonumber \\ &=&  \frac{T}{d}\bra{\psi_{QI}\left(\tau\right)}\left(H_{S} +H_{I} -H\right)\ket{\Psi} + \ket{\varepsilon_{s}} \nonumber \\ &=&  H_{S}\frac{T}{d}\ket{\psi_{S}(\tau)}_e + \frac{T}{d}\bra{\psi_{QI}\left(\tau\right)}H_{I}\ket{\Psi} + \ket{\varepsilon_{s}}. \nonumber \er  


Our interest is in obtaining an equation of motion for a mixed system state with a gravitationally induced interaction. Therefore, we proceed by obtaining a closed form for the Schr{\"o}dinger-like equation for the interaction Hamiltonian \be H_I = -\mathds{G} H_S\otimes H_C, \ee where we defined $\mathds{G} \coloneqq  G/c^4x$, \br \bra{\psi_{QI}(\tau)}H_{I}\ket{\Psi} &=& -\mathds{G}H_S\bra{\psi_{QI}\left(\tau\right)}H_{C}\ket{\Psi} \nonumber \\ &=& \mathds{G}H_S\hspace{-0.1cm}\left[i\frac{d}{T}\frac{d}{d\tau}\bra{\psi_{QI}\left(\tau\right)} + \frac{d}{T}i\bra{\varepsilon'}\right]\ket{\Psi} \nonumber \\ &=& i\mathds{G}H_S\hspace{-0.12cm}\left[\frac{d}{T}\sum_{k}\frac{d}{d\tau}{\psi^{*}\left(\tau; k\right)}\braket{\theta_k|\Psi} + \frac{d}{T}i\braket{\varepsilon'|\Psi}\right] \nonumber \\ &=& i\mathds{G}H_S\hspace{-0.12cm}\left[\frac{d}{T\sqrt{d}}\sum_{k}\frac{d}{d\tau}{\psi^{*}\left(\tau; k\right)}\ket{\psi_{S}(k)} + \frac{d}{T}\ket{\varepsilon_{s}}\right], \nonumber \er hence, \be \bra{\psi_{QI}\left(\tau\right)}H_{I}\ket{\Psi} = i\mathds{G}H_S\frac{d}{T}\frac{d}{d\tau}\ket{\psi_{S}(\tau)}_e + \mathds{G}H_S\frac{d}{T}\ket{\varepsilon_{s}}. \ee With this result we go back to Eq.(\ref{seg}), getting

\begin{widetext} \br\label{quas} i\frac{d}{d\tau}\ket{\psi_{S}(\tau)} &=& H_{S}\frac{T}{d}\ket{\psi_{S}(\tau)}_e + i\mathds{G}H_S\frac{d}{d\tau}\ket{\psi_{S}(\tau)}_e + \mathds{G}H_S\frac{d}{T}\ket{\varepsilon_{s}} + \ket{\varepsilon_{s}} \nonumber \\  &=& H_{S}\frac{T}{d}\ket{\psi_{S}(\tau)}_e + \mathds{G}H_S\left[H_{S}\frac{T}{d}\ket{\psi_{S}(\tau)}_e + i\mathds{G}H_S\frac{d}{d\tau}\ket{\psi_{S}(\tau)}_e + \mathds{G}H_S\frac{d}{T}\ket{\varepsilon_{s}} + \ket{\varepsilon_{s}} \right] + \mathds{G}H_S\frac{d}{T}\ket{\varepsilon_{s}} + \ket{\varepsilon_{s}} \nonumber \\  &=& H_{S}\frac{T}{d}\ket{\psi_{S}(\tau)}_e + \mathds{G}H^{2}_S\frac{T}{d}\ket{\psi_{S}(\tau)}_e + \left(\mathds{G}H_S + \mathds{G}H_S\frac{d}{T} + 1 \right)\ket{\varepsilon_{s}} + \mathcal{O}(\mathds{G}^2) \nonumber \\  &=& H_{S}\frac{T}{d}\ket{\psi_{S}(\tau)}_e + \mathds{G}H^{2}_S\frac{T}{d}\ket{\psi_{S}(\tau)}_e + \ket{\varepsilon_{sg}} + \mathcal{O}(\mathds{G}^2), \er \end{widetext}

The error will be exponentially small in dimension if the Hamiltonian of the system $H_S$ is bounded and that the system state in relation to the time basis is normalized, i.e., $\braket{\psi_{S}(k)|\psi_{S}(k)}=1$. Being both reasonable demands, therefore, as discussed previously, with an appropriate dimension size we can neglect it. Disregarding terms of second order and above of $\mathds{G}$, we obtain our approximate Schr{\"o}dinger equation \be \label{oursch} i\frac{d}{d\tau}\ket{\psi_{S}(\tau)}_e= H_{S}\frac{T}{d}\ket{\psi_{S}(\tau)}_e + \mathds{G}H^{2}_S\frac{T}{d}\ket{\psi_{S}(\tau)}_e. \ee 

Now we consider a mixture of the pure system states in Eq.(\ref{sever}) to be \be \rho_{S}(\tau) =\sum_\iota p_\iota \ket{\psi^{\iota}_{S}(\tau)}_e\bra{\psi^{\iota}_{S}(\tau)},\ee where $\rho_{S}(\tau) = U(\tau)\rho_{S}(0)U^{\dagger}(\tau)=\sum_\iota p_\iota \ket{\psi^{\iota}_{S}(\tau)}_e\bra{\psi^{\iota}_{S}(\tau)}$. By taking the derivative of this state and using Eq.(\ref{oursch}) we find \be \label{ourvon} \frac{d\rho_{S}(\tau)}{d\tau} = -i\frac{T}{d}[H_{S},\rho_{S}(\tau)] -i\mathds{G}\frac{T}{d}[H^{2}_S,\rho_{S}(\tau)]. \ee We readily see that the effect of the gravitational interaction appears in the form of a second term with the system Hamiltonian squared, analogously to the Schr{\"o}dinger-like equation. Therefore we can consider the effect of the Hamiltonian $H_{d}= \frac{T}{d}H_S\left(\mathds{1} + \mathds{G}H_S\right)$, which seems to be analogous to a time dilation effect for quantum clocks in the presence of gravity \cite{cas}. We note that here we only considered the interaction mediated by gravity without explicitly introducing time dilation. Then, we see that the interaction between clock and system state does not induce a non-unitary behavior in the evolution of the system state, provided that Eq.(\ref{oursch}) holds. Therefore, as in the ideal clock case (see Appendix C) there is no decoherence. 

To obtain Eq.(\ref{ourvon}), the evolution is considered with negligible clock error. Now we continue by considering the effect of the small contribution given by the error of the QIC. Let us use the following transformations  \be \tilde{\rho}(\tau) = e^{iH_{d}\tau}\rho_{S}(\tau)e^{-iH_{d}\tau}\ee and \be \tilde{V}(\tau) = e^{iH_{d}\tau}V(\tau)e^{-iH_{d}\tau},\ee where we defined the potential $V\equiv \frac{\mathds{G}_H \varepsilon'}{\sqrt{d}}\ket{\varepsilon_{sg}} {_e\hspace{-0.07cm}}\bra{\psi_{S}(\tau)}$ (more details can be found in Appendix D). With those definitions, the initial state of the system will coincide with the transformed initial state $\tilde{\rho}(0)=\rho_S(0)$, and we obtain \be \frac{d\tilde{\rho}(\tau)}{d\tau} = -i[\tilde{V}(\tau), \tilde{\rho}(\tau)]. \ee Integrating the above equation and iterating it we get \be\label{firt2} \frac{d\tilde{\rho}(\tau)}{d\tau} = -i[\tilde{V}(\tau), \tilde{\rho}(0)] -\int^{\tau}_{0}{[\tilde{V}(\tau), [\tilde{V}(s), \tilde{\rho}(s)]]ds} \ee with this we get an equation that is only dependent on the density operator of the system of interest. We assume weak coupling between clock and system due to the nature of the assumed potential, i.e., gravitational, and expand the transformed density operator around $\tau$, i.e., \be\tilde{\rho}(s) = \tilde{\rho}(\tau) + (s-\tau)\frac{d\tilde{\rho}(\tau)}{d\tau} + \mathcal{O}((s-\tau)^{2}).\ee It, it is easy to see that the derivatives will contribute with higher orders terms, retaining only terms up to second order we reach  \be \frac{d\tilde{\rho}(\tau)}{d\tau} = -i[\tilde{V}(\tau), \rho(0)] -\int^{\tau}_{0}{[\tilde{V}(\tau), [\tilde{V}(s), \tilde{\rho}(\tau)]]ds}. \ee Undoing the transformation, we find  \br \frac{d\rho_S(\tau)}{d\tau} &=& -i[H_{d},\rho_S(\tau)] -i[V(\tau),U^{\dagger}\rho_S(0)U] \nonumber \\ && - \int^{\tau}_{0}{[V(\tau), [V(s), \rho(\tau)]_{\tau-s}]ds},\er where \be[V(s), \rho(\tau)]_{t-s} = e^{-iH_{d}(\tau-s)}[V(s), \rho(\tau)]e^{iH_{d}(\tau-s)}. \nonumber \ee Finally, making use of the Baker-Campbell-Hausdorff formula \cite{lie}, \be U^{\dagger}\rho_{S}(0)U = \rho_S(0) + i\tau [H_{d},\rho_S(0)] - \frac{\tau^2}{2}[H_{d},[H_{d},\rho_S(0)]] + \mathcal{O}(\tau^{3}), \nonumber \ee to expand the second term of the equation above. For a sufficiently small time step $\tau$, we may truncate this series into first order in $\tau$, finding \begin{widetext}\be\label{our} \frac{d\rho_S(\tau)}{d\tau} = -i[H_{d},\rho_S(\tau)] -i[V(\tau),\rho_S(0)] + \tau[V(\tau),[H_{d},\rho_S(0)]]  - \int^{\tau}_{0}[V(\tau), [V(s), \rho(\tau)]_{\tau-s}]ds \hspace{0.3cm}\ee \end{widetext} This is the equation of motion for the mixed state conditioned on clock time. We can see that apart the first term, which describes unitary evolution with a dilated time  Hamiltonian, as seen in Eq.(\ref{oursch}), we obtain a fourth term associated with decoherence of the system state. The potential constructed, can also be written as \be V(\tau) = \frac{\mathds{G}_H \varepsilon'}{\sqrt{d}}V_{k,k'}(\tau). \ee If we expand this potential coefficient in order to show the influence of the error and gravitational constant we get\begin{align} V(\tau) &\propto \left(1+\mathds{G}H_S + \mathds{G}H_S\frac{d}{T}\right)^2 \nonumber \\ & = \left[1+\mathds{G}H_S\left(1+\frac{d}{T}\right) + \left[\mathds{G}H_S\left(1+\frac{d}{T}\right)\right]^2\right]. \end{align} For this potential, the energy given by the Hamiltonian is going to be the energy of the system state $\ket{\psi_S(k)}$, the system state related to the time basis. Then, the decoherence term will have contributions of the order of energy and squared energy,  differently to the equation of motion obtained in Ref.\cite{cas}, which has a dependence only in respect to the square of energy. We also note that in the case that the energy $\ket{\psi_S(k)}$ is negligible we still have decoherence, due to the non-ideal nature of the clock. Moreover, in Eq.(\ref{our}) there are two new terms, the second and the third ones, both dependent on the initial system state $\rho_S(0)$. Given the dependence on initial conditions this shows that this equation of motion is non-linear in nature, differently from other results, in the context of gravitation interaction \cite{cas,anas,casl2}. 

We should emphasize that for our case, the non-unitary and non-linear effects arise due to the non-ideal nature of the chosen clock model, and the ``idealness" can be related to its dimension, showing the role that the size of the clock plays. The smaller its size the more pronounced these new terms become, while the larger it is, more similar to an ideal clock and therefore less noticeable non-unitary/non-linear behaviour.

\section{Conclusion}

In this work, we further develop the Page-Wootters formulation by examining how the qualities of the clock, here considered as non-idealness, influenced the dynamics of a general mixed system state when there is an interaction between clock and system. To this end, we utilized the quasi-ideal clock, a finite clock that can approximate an ideal clock really well, being continuous for any arbitrarily small time value.

When the considered interaction is intermediate through gravity, we obtain an equation of motion that in the limit of very large dimension, when we can neglect the errors of the quasi-ideal clock, the evolution is unitary, and interestingly, containing a term that appears to indicate a time dilation effect. We note here that we do not consider the states to be in a superposition of proper times.

This case of high enough dimension is akin to considering that the quasi-ideal clock behaves as an ideal clock. Non-unitarity in the equation of motion is achieved when we do account for those errors, finding terms that resemble decoherence. We also obtained two additional terms dependent on the system state's initial conditions, making the equation to be non-linear. These two additional terms are distinct, one resembling unitary dynamics in relation to the perturbation and the other as decoherence involving the time dilated Hamiltonian. It is noteworthy that non-unitarity here can be connected to the dimension of our finite clock. Meanwhile, an infinite dimension will be akin to have an ideal clock not having a large enough dimension will contribute to decoherence of the clock and attenuated non-unitary effects on the system state.

\section*{Acknowledgements}
The authors would like to acknowledge financial support from Brazilian funding agencies CNPq (Grants No. 142350/2017-6, 307028/2019-4), FAPESP (Grant No. 2017/03727-0) and the Brazilian National Institute of Science and Technology of Quantum Information (INCT/IQ) [CNPq INCT-IQ (465469/2014-0)].

\bibliographystyle{apsrev4-1}

\begin{thebibliography}{100}
\bibitem{wood1} S. Khandelwal, M. P. Lock,   and M. P. Woods, Quantum 4, 309 (2020). 
\bibitem{smith1} A. R. H. Smith, M. Ahmadi, Quantum clocks observe classical and quantum time dilation, Nat. Commun. 11,5360 (2020).
\bibitem{bose} S. Bose, A. Mazumdar, G. W. Morley, H. Ulbricht, M. Toro\v{s}, M. Paternostro, A. A. Geraci, P. F. Barker, M. S. Kim, G. Milburn, Spin Entanglement Witness for Quantum Gravity, Phys. Rev. Lett. 119, 240401 (2017).
\bibitem{chia} C. Marletto and V. Vedral, Gravitationally Induced Entanglement between Two Massive Particles is Sufficient Evidence of Quantum Effects in Gravity, Phys. Rev. Lett. 119, 240402 (2017).
\bibitem{carlo} F. Hellmann, M. Mondragon, A. Perez, C. Rovelli, Multiple-event probability in general-relativistic quantum mechanics, Phys. Rev. D 75, 084033 (2007).
\bibitem{paw} D. N. Page, W. K. Wootters, Evolution without evolution: dynamics described by stationary observables. Phys. Rev. D 21, 2885 (1983).
\bibitem{f3} W. K. Wootters, ``Time" replaced by quantum correlations, Int. J. Theor. Phys. 23, 701 (1984).
\bibitem{diaz} N. L. Diaz, J. M. Matera, R. Rossignoli, History state formalism for scalar particles, Phys. Rev. D 100, 125020 (2019).
\bibitem{foti} C. Foti, A. Coppo, G. Barni, A. Cuccoli, P. Verrucchi, Time and classical equations of motion from quantum entanglement via the Page and Wootters mechanism with generalized coherent states, Nat. Commun. 12, 1787 (2021).
\bibitem{gamb1} R. Gambini, R. A. Porto, J. Pullin, S. Torterolo, Conditional probabilities with Dirac observables and the problem of time in quantum gravity, Phys. Rev. D 79, 041501(R) (2009).
\bibitem{hoen} P. A. Hoehn, A. R. H. Smith, M. P. E. Lock, The Trinity of Relational Quantum Dynamics, arXiv:1912.00033 (2019).
\bibitem{pegg} D. T. Pegg, Time in a quantum mechanical world, J. Phys. A 24, 3031 (1991).
\bibitem{f} P. A. H\"{o}hn, A. Vanrietvelde, How to switch between relational quantum clocks, New J. Phys. 22 123048 (2020).
\bibitem{f1} R. S. Carmo, D. O. Soares-Pinto, Quantifying resources for the Page-Wootters mechanism: Shared asymmetry as relative entropy of entanglement, Phys. Rev. A 103, 052420 (2021).
\bibitem{f2} A. Boette, R. Rossignoli, N. Gigena, M. Cerezo, System-time entanglement in a discrete-time model, Phys. Rev. A 93, 062127 (2016).
\bibitem{f4} I. L. Paiva, M. Nowakowski, E. Cohen, 
Nonlocality in quantum time via modular operators, arXiv:2104.09321 (2021).
\bibitem{f5} I. L. Paiva, A. C. Lobo, E. Cohen, Flow of time during energy measurements and the resulting time-energy uncertainty relations, arXiv:2106.00523 (2021).
\bibitem{glm} V. Giovannetti, S. Lloyd, and L. Maccone. Quantum Time. Phys. Rev. D 79, 945933 (2015).
\bibitem{casl} E. Castro-Ruiz,  F. Giacomini,  A. Belenchia, \v{C}. Brukner, Quantum clocks and the temporal localisability of events in the presence of gravitating quantum systems, Nat. Commun. 11, 2672 (2020).
\bibitem{smith3} A. R. H. Smith, M. Ahmadi., Quantum clocks observe classical and quantum time dilation, Nat. Commun. 11, 5360 (2020).
\bibitem{mischa}  M. P. Woods, R. Silva, J. Oppenheim, Autonomous Quantum machines and finite-sized clocks, Ann. Henri Poincar\'e 20, 125 (2018).
\bibitem{cont} M.  P.  Woods, A.  M.  Alhambra, Continuous groups of transversal gates for quantum error correcting codes from finite clock reference frames, Quantum 4,  245 (2020).
\bibitem{smith2} A. R. H. Smith, M. Ahmadi, Quantizing time: interacting clocks and systems. Quantum 3, 160 (2019).
\bibitem{whe} J. A. Wheeler, in Battelle Rencontres: 1967 Lectures on Mathematical Physics (Benjamin, New York, 1968).
\bibitem{wit} B. S. DeWitt, Quantum Theory of Gravity. I. The Canonical Theory. Phys. Rev. 160, 1113 (1967).
\bibitem{pash} T. Pashby, Time and the foundations of quantum mechanics, Ph.D. thesis, Faculty of the Dietrich School of Arts and Sciences, University of Pittsburgh, Pittsburgh, 2014.
\bibitem{sw} H. Salecker, E.P. Wigner, Quantum limitations of the measurement of spacetime distances, Phys. Rev. 109, 571 (1958).
\bibitem{swp}  A. Peres, Measurement of time by quantum clocks, Am. J. Phys. 48, 552 (1980).
\bibitem{roto} M. Cal{\c c}ada, J. T. Lunardi, L. A. Manzoni, On the Salecker-Wigner-Peres clock and double barrier tunneling, Phys. Rev. A 79 012110 (2009).
\bibitem{cas}  E. Castro Ruiz, F. Giacomini, \v{C}. Brukner, Entanglement of quantum clocks through gravity. Proc. Natl. Acad. Sci. USA 114, E2303 (2017).
\bibitem{lie} V. S. Varadarajan. Lie groups, Lie algebras, and their representations. Springer-Verlag New York, (1984).
\bibitem{anas} C. Anastopoulos, B. L. Hu, A master equation for gravitational decoherence: probing the textures of spacetime, Class. Quantum Grav. 30, 165007 (2013).
\bibitem{casl2} I. Pikovski, M. Zych, F. Costa, \v{C}. Brukner, Universal decoherence due to gravitational time dilation, Nat. Phys. 11, 668 (2015).
\bibitem{fava} T. Favalli, A. Smerzi, Time observables in a timeless universe, Quantum 4, 354 (2020). 
\end{thebibliography}

\begin{widetext}

\section*{Appendix A: Derivation of the global state}

We can show that the natural choice for the global state used in Eq.(\ref{sever}) is connected, to the global state using the SWP clock, which is has essentially the same construction that was provided in Ref.\cite{fava}. Here we will repeat this construction with a more appropriate notation and then present the derivation for our global state.

We start with a generic state for the universe \be \ket{\Psi} = \sum^{d-1}_{n=0}\sum^{d_S-1}_{m=0} p_{n,m} \ket{E_n}\ket{E_m}, \ee that can be written as \br \ket{\Psi} &=& \sum^{d_S-1}_{m=0} \tilde{p}_{m} \ket{E= -E_m}\ket{E_m} \nonumber \\ &=& \sum^{d_S-1}_{m=0}\sum^{d-1}_{k=0} \tilde{p}_{m} \ket{\theta_k}\braket{\theta_k|E= -E_m}\ket{E_m} \nonumber \\ &=& \frac{1}{\sqrt{d}}\sum^{d_S-1}_{m=0}\sum^{d-1}_{k=0} \tilde{p}_{m}e^{-i2\pi jk/d} \ket{\theta_k}\ket{E_m} \nonumber \\ &=& \frac{1}{\sqrt{d}}\sum^{d-1}_{k=0} \ket{\theta_k}\ket{\psi_S(k)}, \er where in the second line we used the resolution of the identity for time states. 
In order to reach our result we use the covariant POVM generated by the Quasi-Ideal states \cite{cont}, that in the limit of large dimension is approximately $\{P_{QI}(\tau) \coloneqq U(\tau)\ket{\psi_{QI}(0)}\bra{\psi_{QI}(0)}U^{\dagger}(\tau)\}_{\tau \in [0,T]}$, hence , \br \ket{\Psi} &=& \frac{1}{\sqrt{d}}\sum^{d-1}_{k=0} \frac{1}{T}\int^{T}_{0} d\tau \ket{\psi_{QI}(\tau)}\braket{\psi_{QI}(\tau)|\theta_k}\ket{\psi_S(k)} \nonumber \\ &=& \frac{1}{\sqrt{d}T}\sum^{d-1}_{k=0}\int^{T}_{0} d\tau \ket{\psi_{QI}(\tau)}\psi^*(\tau; k)\ket{\psi_S(k)} \nonumber \\ &=& \frac{1}{T}\int^{T}_{0} d\tau \ket{\psi_{QI}(\tau)} \frac{1}{\sqrt{d}}\sum^{d-1}_{k=0}\psi^*(\tau; k)\ket{\psi_S(k)}, \er using the definition given in Eq.(\ref{def}) we get \be \ket{\Psi}= \frac{1}{T}\int^{T}_{0} d\tau \ket{\psi_{QI}(\tau)}\ket{\psi_S(\tau)}.\ee 

\section*{Appendix B: Time evolution of the QIC}

Here we give a proof for Eq.(\ref{sc}), the idea is to use Lemma 8.01 of Ref.\cite{mischa} to obtain the infinitesimal evolution of the QIC, and then employ the definition of derivative \br \frac{d}{d\tau}\ket{\psi_{QI}(\tau)} &=& \sum_{k}\frac{d}{d\tau}\psi(\tau;k)\ket{\theta_k} \nonumber \\ &=& \sum_{k}\left[\lim_{\delta \rightarrow 0}\frac{\psi(\tau+\delta;k)-\psi(\tau;k)}{\delta}\right]\ket{\theta_k} \nonumber \\ &=& \lim_{\delta \rightarrow 0}\frac{\sum_{k}\psi(\tau+\delta;k)\ket{\theta_k}-\sum_{k}\psi(\tau;k)\ket{\theta_k}}{\delta} \nonumber \\ &=& \lim_{\delta \rightarrow 0}\frac{\ket{\psi_{QI}(\tau)} - i\delta (T/d) H_C\ket{\psi_{QI}(\tau)}-\ket{\varepsilon}_i-\ket{\psi_{QI}(\tau)}}{\delta} \nonumber \\ &=& - \frac{iTH_C}{d}\ket{\psi_{QI}(\tau)} - \lim_{\delta \rightarrow 0}\frac{\ket{\varepsilon}_i}{\delta}. \er  The error is will be something like $ \ket{\varepsilon}_i = \sum_{k\in S_d(k_0)}\left[\delta(\varepsilon_1+\varepsilon_2+\varepsilon_3) + \delta^2 C\right]\ket{\theta_k}$, where the form of each error can be found in Ref.\cite{mischa}, therefore its limit will be \br \lim_{\delta \rightarrow 0}\frac{\ket{\varepsilon}_i}{\delta} &=& \lim_{\delta \rightarrow 0}\sum_{k\in S_d(k_0)}\frac{\left[\delta(\varepsilon_1+\varepsilon_2+\varepsilon_3) + \delta^2 C\right]}{\delta}\ket{\theta_k} \nonumber \\ &=& \lim_{\delta \rightarrow 0}\sum_{k\in S_d(k_0)}\left[(\varepsilon_1+\varepsilon_2+\varepsilon_3) + \delta C\right]\ket{\theta_k} \nonumber \\ &=& \sum_{k\in S_d(k_0)}(\varepsilon_1+\varepsilon_2+\varepsilon_3)\ket{\theta_k} \nonumber \\ &=& \sum_{k\in S_d(k_0)}\varepsilon'\ket{\theta_k} = \ket{\varepsilon'}.\er Hence, \be |\hspace{-0.05cm}|\ket{\varepsilon'}|\hspace{-0.05cm}| \leq 
    \begin{cases}
      2\pi A \left(2\sqrt{d}\left(\frac{1}{2}+\frac{1}{2\pi d}+\frac{1}{1-e^{\pi}}\right)e^{\frac{\pi d}{4}} + \frac{1}{2}+\frac{1}{2\pi d}+ \frac{1}{1-e^{\pi}}\right)e^{\frac{\pi d}{4}}& \text{if $\sigma = \sqrt{d}$}\\
			 2\pi A \left(2\sigma\left(\frac{\alpha_0}{2}+\frac{1}{2\pi \sigma^2}+\frac{1}{1-e^{\pi \sigma^2\alpha_0}}\right)e^{\frac{\pi \sigma^2\alpha_0}{4}} + \left(\frac{1}{2\pi d}+\frac{d}{2\sigma^2} + \frac{1}{1-e^{\frac{\pi d}{\sigma^2}}} + \frac{1}{1-e^{\frac{\pi d^2}{\sigma^2}}} \right)e^{\frac{\pi d^2}{4\sigma^2}}\right) & \text{otherwise}
    \end{cases},  \ee which also decreases exponentially with dimension, where $\alpha_0 \in (0,1]$ is a parameter used to quantify how close $j_0$ is from the edge of the energy spectrum \cite{mischa}.

\section*{Appendix C: Equation of motion for the ideal clock}

Following the normal prescription to generalize Eq.(\ref{amd}), taking the state of the system to be $\ket{\psi_{S}(\tau)}$ we get \be \rho_{S}(0) = \sum_\iota p_\iota \ket{\psi^{\iota}_{S}(0)}\bra{\psi^{\iota}_{S}(0)},\ee with $\rho_{S}(\tau) = U(\tau)\rho_{S}(0)U^{\dagger}(\tau)=\sum_\iota p_\iota \ket{\psi^{\iota}_{S}(\tau)}\bra{\psi^{\iota}_{S}(\tau)}$, then

 \br \frac{d\rho_{S}(\tau)}{d\tau} &=& \sum_\iota p_\iota \left(\frac{d}{d\tau}\{\ket{\psi^{\iota}_{S}(\tau)}\}\bra{\psi^{\iota}_{S}(\tau)} + \ket{\psi^{\iota}_{S}(\tau)}\frac{d}{d\tau}\{\bra{\psi^{\iota}_{S}(\tau)}\} \right) \nonumber \\ &=& -i\sum_\iota p_\iota \left(\{H_S \ket{\psi^{\iota}_S(\tau)} + H_k \ket{\psi^{\iota}_S(\tau)}\}\bra{\psi^{\iota}_{S}(\tau)} - \ket{\psi^{\iota}_{S}(\tau)}\{\bra{\psi^{\iota}_S(\tau)}H_S  + \bra{\psi^{\iota}_S(\tau)}H_k\} \right),\er to proceed we first need to obtain the action of the integral operator $H_{k}$. Here we consider a gravitational interaction between clock and system, therefore the Hamiltonian of the universe will be \be H = H_S + H_C - \frac{G}{c^{4}x}H_S\otimes H_C,\ee where $G$ is the gravitational constant, $x$ is the distance between clock and system and $c$ is the speed of light. We are using the idealized momentum clock $H_C = P_C$, then, the kernel $K(\tau,\tau')$ can be computed as follows \br K(\tau,\tau') &=& -\frac{G}{c^4d}H_S\bra{\tau}P_C\ket{\tau'} \nonumber \\ &=& -\frac{G}{c^4d}H_S\left[ \int{dp p \braket{\tau|p}\braket{p|\tau'}} \right] \nonumber \\ &=& -\frac{G}{c^4d}H_S\left[ \frac{1}{2\pi} \int{dp p e^{-ip(\tau'-\tau)}} \right] \nonumber \\ &=& -\frac{G}{c^4d}H_S i\dot{\delta}(\tau'-\tau),\er where the dot on the delta was used in the place of the time derivative. The adjoint kernel $K^{\dagger}(\tau,\tau')$ is readily obtained. With this \br H_{k}\ket{\psi_{S}(\tau)} &=& -\int \frac{G}{c^4d}H_S i\dot{\delta}(\tau'-\tau)\ket{\psi_{S}(\tau')}d\tau' \nonumber \\ &=& i\frac{G}{c^4d}H_S\int \delta(\tau'-\tau)\frac{d}{d\tau'}\ket{\psi_{S}(\tau')}d\tau' \nonumber \\ &=& i\frac{G}{c^4d}H_S\frac{d}{d\tau}\ket{\psi_{S}(\tau)}\er and its adjoint \br \bra{\psi_{S}(\tau)}H_{k} &=& i\frac{G}{c^4d}\int \dot{\delta}(\tau'-\tau)\bra{\psi_{S}(\tau')}H_Sd\tau' \nonumber \\ &=& -i\frac{G}{c^4d}\int \delta(\tau'-\tau)\frac{d}{d\tau'}\bra{\psi_{S}(\tau')}H_S d\tau' \nonumber \\ &=& -i\frac{G}{c^4d}H_S\frac{d}{d\tau}\bra{\psi_{S}(\tau)}H_S.\er Then, defining $\mathds{G} := \frac{G}{c^4d}$, we get \br \frac{d\rho_{S}(\tau)}{d\tau} &=& -i\sum_\iota p_\iota \left(\{H_S \ket{\psi^{\iota}_S(\tau)} + H_k \ket{\psi^{\iota}_S(\tau)}\}\bra{\psi^{\iota}_{S}(\tau)} - \ket{\psi^{\iota}_{S}(\tau)}\{\bra{\psi^{\iota}_S(\tau)}H_S  + \bra{\psi^{\iota}_S(\tau)}H_k\} \right) \nonumber \\ &=& -i\sum_{\iota} p_{\iota}\left([H_{S},\ket{\psi^{\iota}_S(\tau)}\bra{\psi^{\iota}_S(\tau)}] +\mathds{G}\left\{iH_S\frac{d\ket{\psi^{\iota}_{S}(\tau)}}{d\tau}\bra{\psi^{\iota}_{S}(\tau)} + i\ket{\psi^{\iota}_{S}(\tau)}\frac{d\bra{\psi^{\iota}_{S}(\tau)}}{d\tau}H_S\right\}\right) \nonumber \\ &=& -i\sum_{\iota} p_{\iota}\left([H_{S},\ket{\psi^{\iota}_S(\tau)}\bra{\psi^{\iota}_S(\tau)}] + \mathds{G}\left\{H^{2}_S\ket{\psi^{\iota}_{S}(\tau)}\bra{\psi^{\iota}_{S}(\tau)} + H_S H_{k}\ket{\psi^{\iota}_{S}(\tau)}\bra{\psi^{\iota}_{S}(\tau)} \right. \right. \nonumber \\ & & \hspace{6cm} \left. \left. -\ket{\psi^{\iota}_{S}(\tau)}\bra{\psi^{\iota}_{S}(\tau)}H^{2}_S - \ket{\psi^{\iota}_{S}(\tau)}\bra{\psi^{\iota}_{S}(\tau)}H_{k}H_S\right\}\right) \nonumber \\ &=& -i\sum_{\iota} p_{\iota}\left([H_{S},\ket{\psi^{\iota}_S(\tau)}\bra{\psi^{\iota}_S(\tau)}] + \mathds{G}[H^{2}_S,\ket{\psi^{\iota}_{S}(\tau)}\bra{\psi^{\iota}_{S}(\tau)}] + \mathcal{O}(\mathds{G}^2)\right),\er where in the second line we used Eq.(\ref{amd}). As all other terms will be of orders superior to $\mathds{G}$ which is proportional to the inverse of $c^4$, we assume these terms to be negligible, therefore our equation of motion will be approximately \be\label{vonpaw} \frac{d\rho_{S}(\tau)}{d\tau} =-i[H_{S},\rho_S(\tau)] -i\mathds{G}[H^{2}_S,\rho_S(\tau)]. \ee		

\section*{Appendix D: Time dependent potential}

To obtain the potential, we start by defining the normalized version of the error $\ket{\varepsilon_{sg}} = \sqrt{\braket{\varepsilon_{s}|\varepsilon_{s}}}\ket{\varepsilon'_{sg}}$, then, from $\ket\varepsilon_{s}\coloneqq i\braket{\varepsilon'|\Psi}$, we have \br \ket{\varepsilon_{sg}} &\coloneqq&  \left(1+\mathds{G}H_S + \mathds{G}H_S\frac{d}{T}\right)i\sum_{k\in S_d(0)}\varepsilon'\braket{\theta_k|\Psi} \nonumber \\ &=& \mathds{G}_H\varepsilon'\sum_{k\in S_d(0)}\braket{\theta_k|\Psi} \nonumber \\ &=& \frac{\mathds{G}_H\varepsilon'}{\sqrt{d}}\sum_{k\in S_d(0)}\ket{\psi_S(k)}.\er Then, we can define \br V \coloneqq \sqrt{\braket{\varepsilon_{s}|\varepsilon_{s}}}\ket{\varepsilon_{sg}}{_{e}\hspace{-0.07cm}}\bra{\psi_S(\tau)},\er in a way that Eq.(\ref{quas}), disregarding terms of second order of $\mathds{G}$,  becomes \be i\frac{d}{d\tau}\ket{\psi_{S}(\tau)}_e = H_{S}\frac{T}{d}\ket{\psi_{S}(\tau)}_e + \mathds{G}H^{2}_S\frac{T}{d}\ket{\psi_{S}(\tau)}_e + V\ket{\psi_S(\tau)}_e. \ee

Just as a note, from the equations above we can see that explicit the potential can be written as \begin{align} V &= \frac{\mathds{G}_H\varepsilon'}{\sqrt{d}}\sum_{k,k'}\psi(\tau;k')\ket{\psi_S(k)}\bra{\psi_S(k')} \nonumber \\  &= \frac{\mathds{G}_H\varepsilon'}{\sqrt{d}}V_{k,k'}. \end{align}

\end{widetext}

\end{document}